\documentclass[sigplan,nonacm]{acmart}

\usepackage[]{hyperref}

\usepackage[normalem]{ulem}
\usepackage{textgreek}
\usepackage{textcomp}

\usepackage{multirow}
\usepackage{xspace}
\usepackage{subfig}
\usepackage[font=small,labelfont=bf]{caption}
\usepackage{graphicx}

\usepackage{fancyhdr}
\usepackage{colortbl}
\usepackage{tabulary}
\usepackage{etoolbox}
\usepackage{booktabs}
\usepackage{mathtools}
\usepackage{pifont}
\usepackage{color}
\usepackage{makecell}

\newcommand{\fig}[1]{Figure~\ref{#1}}

\newcommand{\sect}[1]{Section~\ref{#1}}
\newcommand{\tab}[1]{Table~\ref{#1}}

\newcommand{\proposed}[0]{Mamba-X\xspace}

\begin{document}

\title{Mamba-X: An End-to-End  Vision Mamba Accelerator \\for Edge Computing Devices}
\thanks{This is an extended version of our work, which is accepted for publication at the $44^{th}$
International Conference on Computer-Aided Design (ICCAD), $2025$.}

\author{Dongho Yoon}
\affiliation{%
  \institution{KAIST}
  \country{}
}
\email{dongho.yoon@kaist.ac.kr}

\author{Gungyu Lee}
\affiliation{%
  \institution{KAIST}
  \country{}
}
\email{gungyu.lee@kaist.ac.kr}

\author{Jaewon Chang}
\affiliation{%
  \institution{KAIST}
  \country{}
}
\email{jaewon.chang@kaist.ac.kr}

\author{Yunjae Lee}
\affiliation{%
  \institution{KAIST}
  \country{}
}
\email{yunjae408@kaist.ac.kr}

\author{Dongjae Lee}
\affiliation{%
  \institution{KAIST}
  \country{}
}
\email{dongjae.lee@kaist.ac.kr}

\author{Minsoo Rhu}
\affiliation{%
  \institution{KAIST}
  \country{}
}
\email{mrhu@kaist.ac.kr}

\begin{abstract}

Transformers have proven effective in language modeling but are limited by high computational and memory demands that grow quadratically with input sequence length. State space models (SSMs) offer a promising alternative by reducing attention complexity from $O(L^2)$ to $O(L)$ while also lowering overall memory consumption. Vision Mamba adapts the SSM approach for computer vision tasks, achieving lower latency and memory consumption than traditional transformer models. However, deploying Vision Mamba on edge devices is challenging due to its sequential scan operations, which hinder GPU efficiency. We propose \proposed, an end-to-end Vision Mamba accelerator that includes a systolic scan array to maximize parallelism and minimize memory traffic, along with a hybrid, hardware-friendly quantization technique to reduce memory usage and improve hardware efficiency without sacrificing accuracy.

\end{abstract}

\maketitle 
\pagestyle{plain} 

\section{Introduction}

Attention-based transformers~\cite{Attention} have demonstrated their effectiveness in language modeling, widely being employed in various  domains~\cite{vision_transformer, CLIP, LLaVA, OV_CLIP}. However, these models are fundamentally limited by their high computational and memory requirements that scale quadratically with the input sequence length. While prior work attempted to mitigate these challenges—primarily through model compression via quantization~\cite{I-BERT,  SmoothQuant, AWQ, vaqf, autovitacc, PTQ_ViT, PTQ4ViT, Repq_vit} or fine-grained computation of the attention mechanism~\cite{Linformer, Reformer, swin_transformer, vitcod, vitality, heatvit, adaptiv, paged_attention}—they do not address the inherent quadratic growth in resource demands associated with the attention operation.

To address these challenges, \emph{state space models} (SSMs) have recently emerged as a promising alternative to attention-based models. Among these SSMs, Mamba~\cite{mamba} is a state-of-the-art SSM proven to be effective for large language models. Specifically, Mamba reduces the complexity of the attention mechanism from $O(L^2)$ to $O(L)$, where $L$ is the input sequence length. Additionally, Mamba eliminates the need to store the score matrix and the key-value cache~\cite{KV_cache}, which is known to incur high memory consumption in conventional attention-based transformer architectures,  making Mamba well-suited for handling long-context AI tasks.

\begin{figure}[t]
\centering
\includegraphics[width=0.485\textwidth]{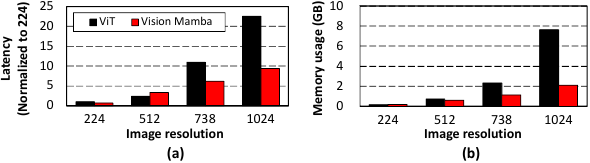}
\caption{Comparison of (a) end-to-end latency and (b) memory consumption when the Vision Transformer (ViT)~\cite{vision_transformer} and Vision Mamba are executed on NVIDIA's Jetson AGX Xavier. As depicted, the advantages of Vision Mamba in terms of compute (latency) and memory efficiency grow as the input image size increases.
}
\label{fig:DeiT_and_Vision_Mamba}
\vspace{-0.5em}
\end{figure}

Inspired by the success of Mamba for language modeling, researchers have explored its application across various workloads, similar to how attention-based models have been extended beyond natural language processing. Vision Mamba~\cite{vision_mamba} is one such representative example, which adapts Mamba for computer vision tasks by extending its architecture from one-dimensional sequential data to two-dimensional image data. As shown in \fig{fig:DeiT_and_Vision_Mamba},  Vision Mamba can efficiently process high-resolution images with significantly lower latency and memory consumption, while still maintaining competitive algorithmic performance compared to traditional transformer-based vision models. 
Driven by the increasing demand for high-resolution computer vision tasks in real-time applications such as autonomous vehicles, smart surveillance, and augmented reality~\cite{high_resolution,gscore,metasapiens, characterizing_edge, blisscam, deepcache}, the importance of efficient vision processing under the tight hardware constraints of edge devices has grown significantly.

Unfortunately, deploying Vision Mamba on resource-limited edge devices presents several challenges. Current transformer-based vision models are generally dominated by highly parallelizable dense GEMM operations, which can be efficiently executed on commodity GPUs. In contrast, Vision Mamba relies on SSM operations that exhibit inherently \emph{sequential} task dependencies. This makes Vision Mamba less suitable for efficient GPU acceleration, leading to significant resource underutilization and suboptimal performance.

Given these challenges, a primary goal of this work is to tackle the computational and memory limitations of Vision Mamba in edge environments. We begin by thoroughly characterizing the deployment of state-of-the-art Vision Mamba on edge GPUs, identifying that \emph{selective scan} operations in the SSM block create a major performance bottleneck. Our key observation is that the inherently sequential nature of the scan operation limits parallelism and introduces significant synchronization overhead, leading to compute underutilization and excessive off-chip memory traffic.

To this end, we propose \proposed, an end-to-end Vision Mamba accelerator that addresses Mamba's computation and memory challenges for edge devices. Key features include a Systolic Scan Array (SSA) architecture based on the Kogge-Stone algorithm, designed to maximize parallelism in the selective scan operation, reduce on-chip buffer usage, and minimize off-chip memory traffic. Additionally, we introduce a hybrid, hardware-friendly quantization technique, tailored to Mamba's data distribution, that mitigates accuracy loss, reduces memory usage, and enhances hardware efficiency with a scaling factor approximation, all without sacrificing model accuracy. Overall, \proposed achieves an average 11.6$\times$ improvement in selective scan throughput, which leads to an average 11.5$\times$ end-to-end energy-efficiency improvement, and 601$\times$ increase in average performance/area, demonstrating its merits for resource-constrained edge devices.

\section{Background}

\subsection{State Space Models and Mamba}

Attention mechanisms have a fundamental limitation in that their computational and memory demands scale quadratically with input data size. To address this, state space models (SSMs) have  emerged as a promising alternative to attentions~\cite{S4, H3, mamba}. Originally developed in scientific fields such as control theory, SSMs efficiently model complex systems using state variables in the continuous-time domain. As illustrated in \fig{fig:ssm}(a), a state variable $state_n$ is recursively updated at each time step $n$ based on the input $u_n$, producing the output $y_n$. Here, $\mathbf{A}$ represents the state transition matrix, while $\mathbf{B}$ and $\mathbf{C}$ serve as input and output transformation matrices, respectively. Notably, $\mathbf{A}$, $\mathbf{B}$, and $\mathbf{C}$ remain constant over time, forming what is known as a linear time-invariant (LTI) system. To implement the LTI system in a digital environment, it must be discretized using a fixed time step ($n$) and step size ($\Delta$), as shown in \fig{fig:ssm}(b). The discrete-time SSM formulation can be derived using the zero-order hold discretization method~\cite{mamba, ZOH}. 

For language models, Mamba~\cite{mamba} adopts selection mechanism into the SSM, called \textit{selective} SSM. While traditional SSMs assume that $\mathbf{A}$, $\mathbf{B}$, $\mathbf{C}$, and $\Delta$ remain fixed across all time steps—limiting their ability to adaptively prioritize important input information—Mamba makes the parameters $\mathbf{B}$, $\mathbf{C}$, and $\Delta$ time-variant, allowing them to be dynamically computed based on the input sequence $u_n$. This enables the model to identify the most relevant input information. Overall, Mamba has proven effective in language modeling as it reduces the computational complexity of the attention mechanism from $O(L^2)$ to $O(L)$ ($L$: input sequence length). By eliminating the need to store the score matrix and key-value cache, both essential in transformer architectures~\cite{Attention, KV_cache}, Mamba significantly reduces memory overhead, making it particularly well-suited for long-context processing.

\subsection{Vision Mamba}
\label{sect:vision_mamba}

In computer vision, remarkable advancements have been made, evolving from early convolutional neural network-based models~\cite{Alexnet, Resnet, Mobilenet} to more sophisticated architectures, including transformer-based models like Vision Transformer (ViT)~\cite{vision_transformer, DETR, CrossViT}. However, the quadratic scaling of computation and memory costs with respect to input image size remains a major challenge with ViT based vision models.

\begin{figure}[t] 
\centering
\includegraphics[width=0.4\textwidth]{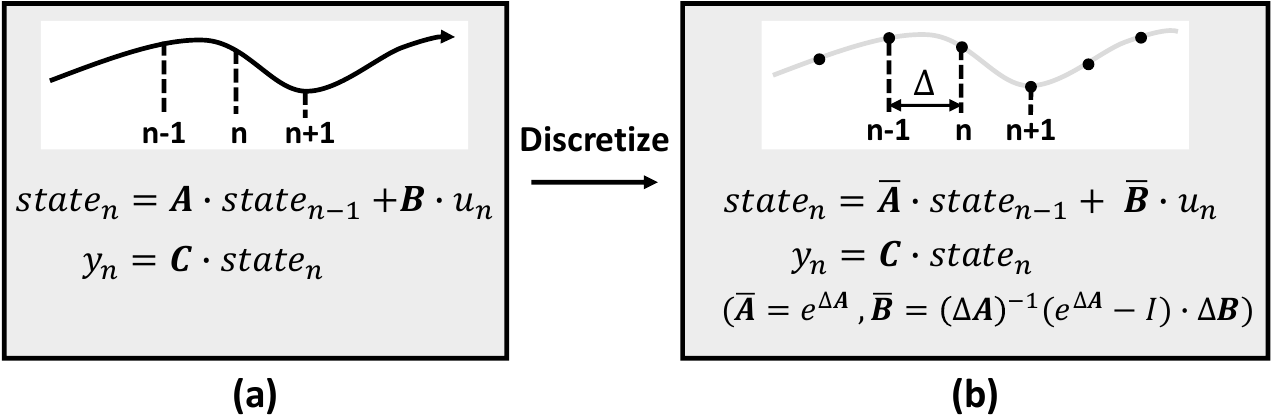}
\vspace{-0.5em}
\caption{SSM in (a) continuous and (b) discrete time domain.}
\label{fig:ssm}
\vspace{-0.5em}
\end{figure}

\begin{figure*}[t] 
\centering
\includegraphics[width=0.93\textwidth]{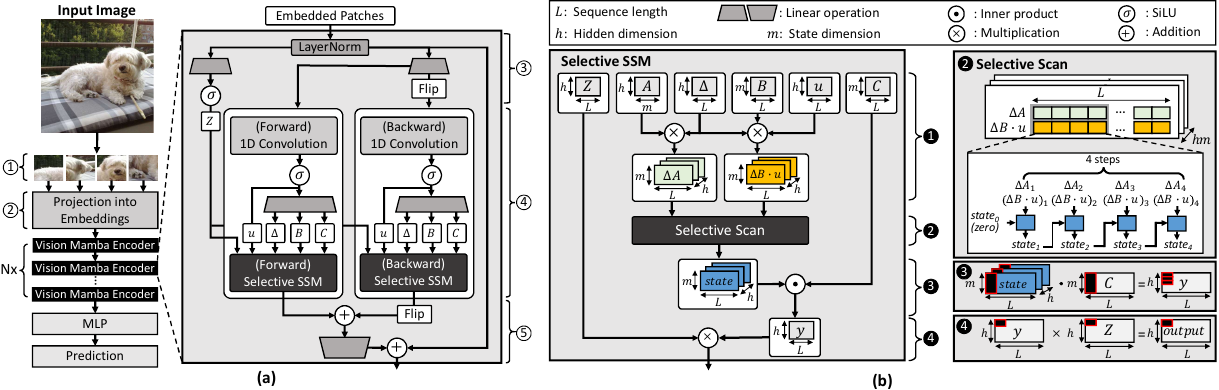}
\caption{(a) Vision Mamba architecture, which substitutes ViT's transformer-based encoders with Vision Mamba encoders. (b) An overview of the selective SSM block. The selective SSM block used in the forward/backward paths of (a) are identical to the selective SSM shown in (b). }
\label{fig:vim_architecture}
\vspace{-0.5em}
\end{figure*}

Inspired by Mamba's success in language modeling, Vision Mamba has recently emerged as a compelling alternative to ViT by integrating the SSM-based Mamba architecture~\cite{mamba} for computer vision tasks. As illustrated in~\fig{fig:vim_architecture}(a), Vision Mamba builds upon the ViT model~\cite{vision_transformer} by replacing traditional transformer-based encoder blocks with Mamba blocks (i.e., the \textit{N} consecutive Vision Mamba encoder blocks). Vision Mamba first partitions the input image into multiple patches, treating each patch as an input token (Step \textcircled{1} in \fig{fig:vim_architecture}(a)). These tokens are then transformed into embeddings (Step \textcircled{2}) and passed into the Vision Mamba encoder. Within each encoder block, the input is first normalized and linearly projected (Step \textcircled{3}). Unlike the original Mamba architecture, which employs a single 1D convolution and the SSM block, Vision Mamba introduces \emph{bidirectional} sequence modeling to effectively capture spatial dependencies in two-dimensional image data. Specifically, it processes input tokens along two spatial directions using forward and backward paths. In each path, every token independently undergoes a 1D convolution and SSM parameter projection (i.e., generating $\Delta$, $\mathbf{B}$, and $\mathbf{C}$), before being processed by the selective SSM block. Here, $\mathbf{Z}$ and $u$ are fed into the selective SSM block along with the SSM parameters (Step \textcircled{4}). Finally, the outputs from both selective SSM blocks are aggregated, projected, and added to the skip connection to produce the final output (Step \textcircled{5}).

\fig{fig:vim_architecture}(b) provides an overview of Vision Mamba's selective SSM block, which replaces the attention layers in conventional transformer-based architectures. In this layer, $\Delta \mathbf{A}$ and $\Delta \mathbf{B} \cdot u$ are first computed using element-wise multiplication (Step \ding{182}). The \textit{selective scan} operation is then performed, which executes across the state dimension ($m$) of both $\Delta\mathbf{A}$ and $\Delta \mathbf{B} \cdot u$. In other words, an $L$-wide vector from the same state dimension of $\Delta\mathbf{A}$ and $\Delta \mathbf{B} \cdot u$ is processed in the selective scan operation. During selective scan, the state variable $state_n$ is recursively computed using $\Delta\mathbf{A}$, $\Delta \mathbf{B} \cdot u$, and the previous state $state_{n-1}$ (see \fig{fig:ssm}(b)). The state variable $state_n$ is sequentially updated across the entire input sequence of length $L$, starting from the initial step ($n=0$). Notably, this computation is executed independently across the hidden dimension ($h$) and the state dimension ($m$) (Step \ding{183}). Finally, the output is computed via the inner product of $\mathbf{C}$ and the selective scan output (Step \ding{184}), followed by element-wise multiplication with $\mathbf{Z}$ (Step \ding{185}).

Overall, Vision Mamba enables efficient high-resolution computer vision processing with lower memory overhead while maintaining competitive performance compared to ViT-based vision models. As the demand for real-time high-resolution computer vision in edge environments grows—and the need for on-device AI increases due to privacy concerns—Vision Mamba's efficiency unlocks new opportunities for vision tasks in edge settings, where hardware constraints require optimized memory and compute usage. However, deploying Vision Mamba on resource-limited edge devices presents several challenges, which we detail in \sect{sect:motivation}.

\subsection{Quantization}
\label{sect:background_quantization}

Quantization compresses models by converting high-precision weights and activations (e.g., FP32) into lower precision (e.g., INT8). A common approach is uniform symmetric quantization, summarized as follows:
\begin{equation}
\label{eq:quant_equation}
s = {{X_{max}}\over{2^{b-1}-1}},\quad X_{q} = \lceil {{X_{f}}\over{s}} \rfloor
\end{equation}

Here \(s\) is the scaling factor, \(X_{max}\) is the maximum value in the quantized tensor, \(b\) is the bit width of the quantized value, and \(X_f, X_q\) are the original floating-point and quantized values, respectively. Dequantization restores \(X_f\) by multiplying \(X_q\) with \(s\), introducing some quantization error. Determining \(X_{max}\) at runtime may incur excessive latency overhead, which motivates the use of \textit{static quantization}~\cite{atom, AWQ}, where scaling factors are pre-computed offline from the activations using calibration samples. 

The selection of optimal quantization granularity is crucial for balancing model accuracy with computational efficiency. In tensor-granularity quantization, a single scaling factor is applied to the entire tensor. In contrast, channel-granularity quantization assigns individual scaling factors to each channel along the hidden dimension~\cite{tender, SmoothQuant, qbert}. The choice between these methods depends on the performance and resource constraints of the deployment platform.

\section{Characterization and Motivation}
\label{sect:motivation}

We use NVIDIA's Jetson AGX Xavier~\cite{JetsonAGXXavier} to identify the key performance-limiting operation in Vision Mamba, analyzing its computational and memory characteristics. 

\subsection{Identifying the Bottleneck in Vision Mamba}
\label{sect:latency_breakdown}

\begin{figure}[t]
\centering
\includegraphics[width=0.485\textwidth]{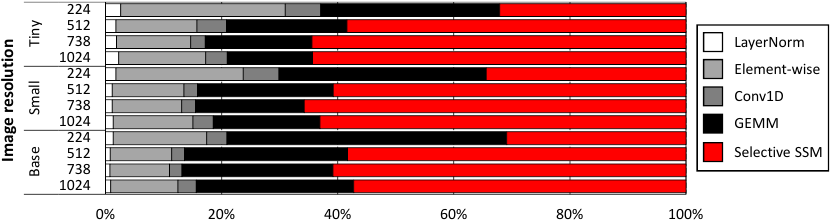}
\caption{Vision Mamba's latency breakdown based on key operations in its encoder block. For images larger than 512$\times $512, selective SSM accounts for up to 60\% of total latency across all models.
}
\label{fig:Latency_breakdown}
\end{figure}

We identify the performance bottlenecks of Vision Mamba by measuring its end-to-end execution time. Our analysis shows that the Vision Mamba encoder block (\fig{fig:vim_architecture}(a)) accounts for the majority of inference time. For example, the 24 encoder blocks in Vision Mamba contribute to 98\% and 99\% of the total latency for image resolutions of 224$\times$224 and 1024$\times$1024, respectively. To this end, \fig{fig:Latency_breakdown} profiles the latency of executing a single Vision Mamba encoder block and breaks it down into key operations: GEMM, LayerNorm, 1D convolution, element-wise operations, and selective SSM. As shown, the selective SSM is the dominant contributor to execution latency. This is because the selective SSM block involves sequential and recursive computations for updating the state variable ($state_n$), which are challenging to efficiently parallelize  on GPUs. In contrast, GEMM operations—typically the most performance-critical in transformer-based models—account for a smaller portion of execution time. This is due to their highly optimized implementation on GPUs, especially as input sizes increase. These results highlight the selective SSM block as a key performance optimization target for edge devices.

\begin{figure}[t]
\centering
\includegraphics[width=0.48\textwidth]{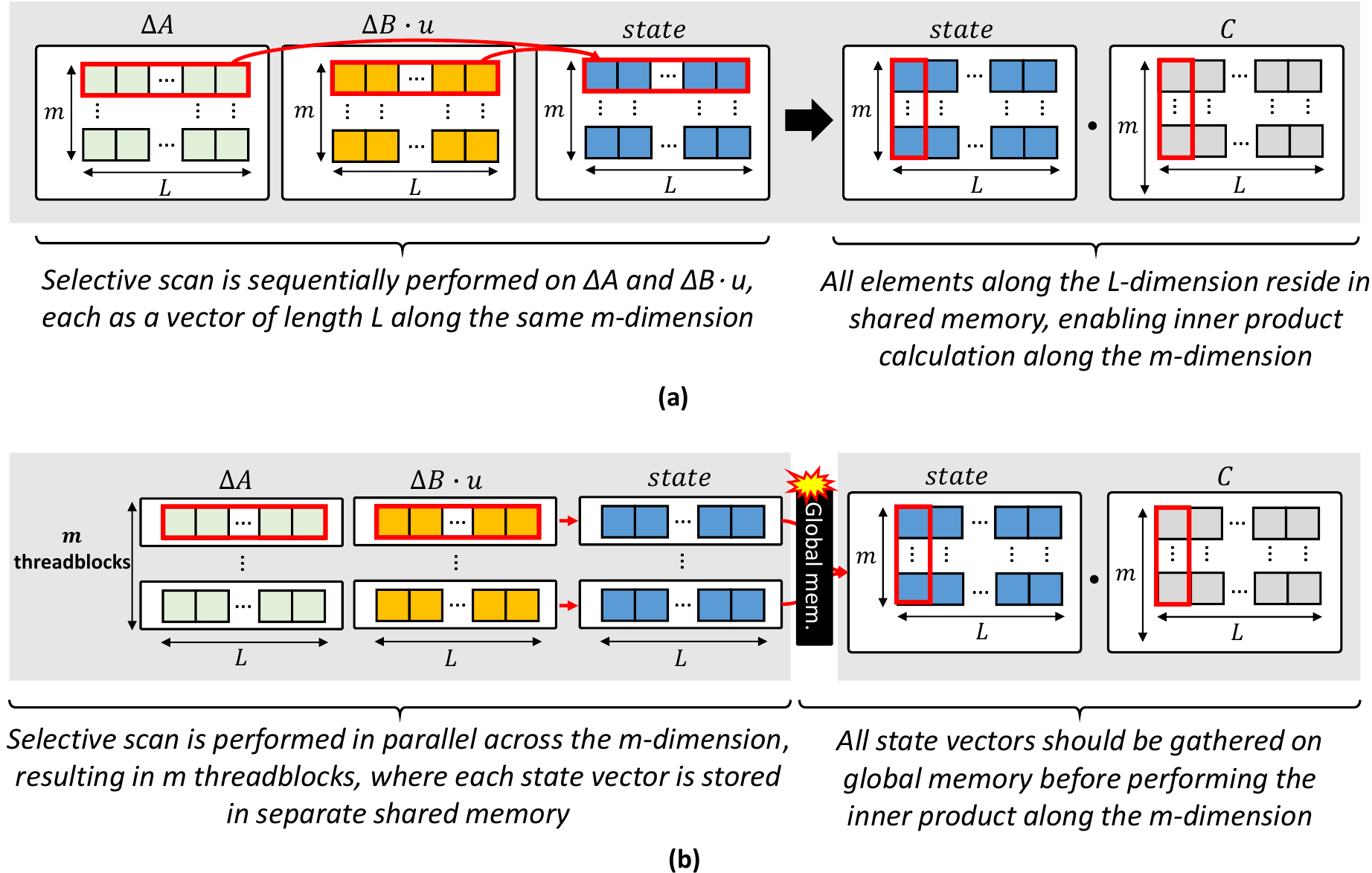}
\caption{Two possible GPU-based implementations of selective SSM operation. (a) The baseline fused selective SSM, which minimizes off-chip traffic at the cost of reduced parallelism. (b) The parallelism-optimized approach helps reap out parallelism in the $m$ dimension at the cost of higher off-chip global memory traffic.}
\label{fig:MambaAccel_Baseline_architecture}
\vspace{-1.0em}
\end{figure}

\subsection{Understanding the Bottleneck in Vision Mamba}
\label{sect:roofline_analysis}

\textbf{Sub-optimal compute utilization due to limited parallelism.} 
The selective scan operation is inherently a sequential computation process along the \emph{L} dimension but opportunities for parallel execution does exist in both the hidden dimension (\emph{h}) and the state dimension (\emph{m}). Nonetheless, the GPU-optimized selective SSM is designed to exploit parallelism only along the hidden dimension (\emph{h}), leaving significant performance untapped. To optimize data movement and minimize off-chip memory accesses, the selective SSM block is implemented as a single, \emph{fused} CUDA kernel (i.e., Steps \ding{182}, \ding{183}, \ding{184}, and \ding{185} in \fig{fig:vim_architecture}(b)). While fusing kernels helps reduce off-chip data movement, it limits the opportunity to exploit available parallelism during the multi-step selective SSM computation.  Specifically, the current design is optimized to maximize data-level parallelism during Step \ding{184} by performing an inner product along the state dimension (\emph{m}). However, under the fused selective SSM implementation, this choice prevents the previous selective scan operation (Step \ding{183}) from fully utilizing parallelism across both the hidden (\emph{h}) and state (\emph{m}) dimensions. The reason parallelism is restricted is illustrated in \fig{fig:MambaAccel_Baseline_architecture}. Performing the inner product along the state dimension (\emph{m}) in Step \ding{184} restricts the CUDA thread block (one which initiates Step \ding{183} and \ding{184} within the fused function call back-to-back) from concurrently scanning rows in parallel along the state dimension (\emph{m}) while executing Step \ding{183}, without leading to significant off-chip memory traffic.

\begin{figure}[t]
\centering
\includegraphics[width=0.48\textwidth]{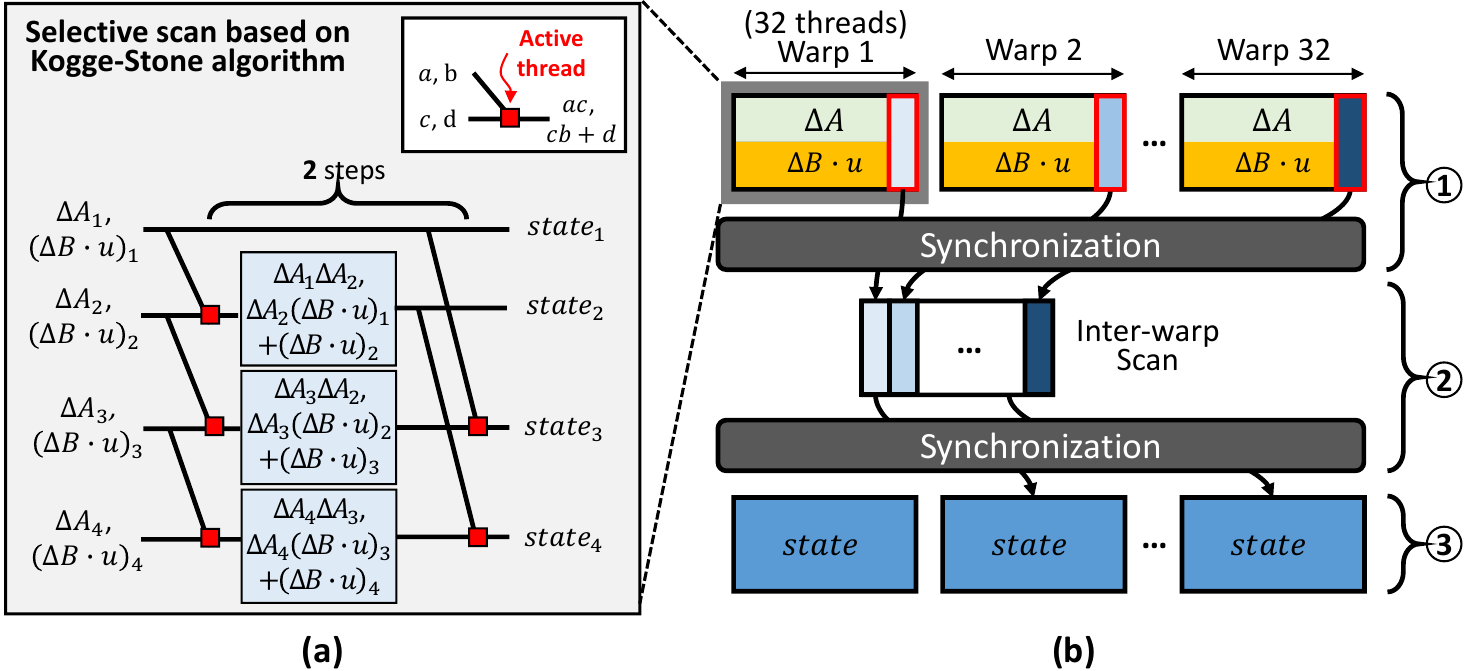}
\caption{(a) Illustration of how the selective scan operation can better leverage parallelism by adopting the Kogge-Stone algorithm where at each step, the result ($cb+d$) is used to compute the state. (b) Execution flow showing selective scan computations across the warps with intermittent inter-warp synchronizations.}
\label{fig:MambaAccel_kogge_stone_implementation}
\vspace{-0.5em}
\end{figure}

\begin{figure}[t]
\centering
\includegraphics[width=0.48\textwidth]{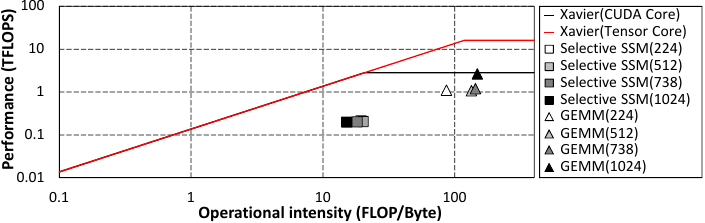}
\caption{Roofline analysis of the selective SSM and GEMM in Vision Mamba on NVIDIA Jetson AGX Xavier. Selective SSM is executed on the CUDA Core, while GEMM is executed on the Tensor Core, unless the cuBLAS runtime opts in on the CUDA core option for better performance (e.g., 512 and 738).}
\label{fig:roofline_analysis}
\vspace{-0.3em}
\end{figure}

To compensate for reduced parallelism, state-of-the-art Vision Mamba~\cite{vision_mamba_github} implementations use the Kogge-Stone algorithm~\cite{kogge-stone} for the selective scan operation (Step \ding{183}). The Kogge-Stone algorithm is a parallel prefix sum (scan) method known for its low depth $O(\log_2 N)$ and high parallelism, making it well-suited for  parallel architectures like GPUs. As shown in~\fig{fig:MambaAccel_kogge_stone_implementation}(a), the Kogge-Stone-based selective scan performs a parallel inclusive scan that efficiently computes the state variable $state_n$, reducing computational complexity to $O(\log_2 N)$. With this algorithm in place, the GPU execution flow proceeds as follows. Each warp performs a local, \emph{intra}-warp selective scan, generating a partial result for its rightmost thread. Multiple warps execute these scans in parallel along the \emph{L} dimension, significantly improving throughput compared to a naive sequential scan operation along the \emph{L} dimension. The warp-level results are then stored in on-chip shared memory and synchronized (Step \textcircled{1} in \fig{fig:MambaAccel_kogge_stone_implementation}(b)). Next, an \emph{inter}-warp selective scan is applied to the warp-level results (Step \textcircled{2}). Finally, each warp retrieves its updated local result from shared memory and applies it to its local elements (Step \textcircled{3}). Although this approach improves parallelism along the \emph{L} dimension, the sequential nature of selective scan introduces synchronization overheads, which becomes problematic as input sequence gets longer. More critically, due to the structure of the Kogge-Stone algorithm, where the number of active computations halves at each step (\fig{fig:MambaAccel_kogge_stone_implementation}(a)), the number of active threads per warp decreases exponentially as the input sequence length $L$ increases. This leads to significant branch divergence~\cite{dynamic_warp_formation, dual_path_execution}, causing substantial compute underutilization.

\fig{fig:roofline_analysis} quantifies the extent of selective SSM's resource underutilization through a roofline analysis of Vision Mamba executed on the baseline GPU system. As shown, the operational intensity and performance of the selective SSM are significantly lower than those of GEMM, whose inefficiency persists regardless of input image size or model scale. Although the selective SSM block is partially parallelized using the Kogge-Stone algorithm (\fig{fig:MambaAccel_kogge_stone_implementation}), its inherent sequential dependencies and significant synchronization overheads make the selective scan operation poorly suited for GPU acceleration, leading to suboptimal performance.

\begin{figure}[t]
\centering
\includegraphics[width=0.47\textwidth]{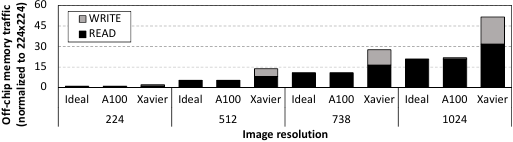}
\caption{Off-chip memory traffic of the selective SSM on NVIDIA A100 and Jetson AGX Xavier, measured using NVIDIA Nsight Compute~\cite{nsight_compute}. All data points are normalized to Ideal's READ when the input image is 224$\times$224. }
\label{fig:Memory_traffic}
\vspace{-1.0em}
\end{figure}

\textbf{High off-chip memory traffic due to limited on-chip shared memory.} The fused selective SSM block is optimized using NVIDIA's CUB~\cite{CUB} library, which allows the SSM block's state variables to be efficiently reused during the scan operation. This approach avoids storing intermediate state variables in off-chip memory by maximizing the usage of on-chip shared memory, providing high efficiency on server-class GPUs such as the NVIDIA A100 featuring large on-chip SRAM capacity. However, edge GPU devices such as the Jetson AGX Xavier have limited on-chip storage, preventing the CUB-based selective SSM block from fully retaining high-reuse intermediate state variables in on-chip shared memory. As a result, overflowing data \emph{spills} to off-chip global memory, leading to excessive off-chip memory traffic and degrading overall performance. 

In \fig{fig:Memory_traffic}, we compare the off-chip memory traffic of the A100 and Jetson AGX Xavier against an oracular, \emph{ideal} GPU design with infinite on-chip storage. On the A100, the off-chip memory traffic closely aligns with the ideal design point, with its traffic size scaling proportionally to the input image size. This indicates that memory accesses on the A100, which contains sufficient on-chip storage, are optimally managed with minimal unnecessary data transfers. In contrast, the Jetson AGX Xavier suffers from significantly higher off-chip traffic, primarily due to the frequent storing and reloading of intermediate data. Unlike the A100, which has ample shared memory to retain intermediate results on-chip, the Jetson AGX Xavier’s limited on-chip shared memory forces frequent spills to off-chip memory. These findings highlight the fundamental challenge of optimizing selective SSM execution for resource-limited edge platforms, where memory constraints play a critical role in overall system efficiency.

\subsection{Motivation}
\label{sect:motivation}
Our characterization revealed that selective SSM is the primary performance bottleneck in Vision Mamba, limiting efficiency in resource-constrained edge devices. Unlike GEMM, which is optimized for parallel execution on GPUs, selective SSM suffers from inherent sequential dependencies, resulting in low compute utilization. Additionally, the limited on-chip SRAM capacity in edge devices makes it challenging to retain high-reuse data on-chip, leading to excessive off-chip memory traffic. These challenges necessitate architectural optimizations to enhance compute efficiency and mitigate memory bottlenecks, motivating our work to improve Vision Mamba’s deployment on resource-limited platforms.

\section{\proposed Architecture and Design}

\begin{figure}[t] 
\centering
\includegraphics[width=0.47\textwidth]{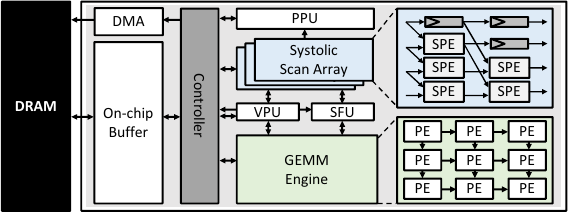}
\caption{High-level overview of our \proposed architecture.}
\label{fig:MambaAccel_Architecture_overview}
\vspace{-0.7em}
\end{figure}

\proposed is an end-to-end Vision Mamba accelerator that synergistically combines a systolic array-style scan array microarchitecture with our hybrid, hardware-friendly quantization scheme tailored for Vision Mamba. In the following subsections, we detail the key facets of our proposal.

\subsection{Architecture Overview}

{\bf Key components.} \fig{fig:MambaAccel_Architecture_overview} provides an overview of \proposed, which contains dedicated compute units for each key operation of Vision Mamba: (1) a DMA unit that orchestrates on-/off-chip data movements, (2) an output-stationary systolic array based GEMM engine~\cite{tpuv1,eyeriss} handling all linear projection operations, (3) a vector processing unit (VPU) conducting all vector operations (e.g., LayerNorm, Conv1D, flip or element-wise operations), (4) a special function unit (SFU) performing non-linear operations within the Vision Mamba model (e.g., SiLU, exponential, and softplus functions), (5) the systolic scan array (SSA) which conducts the selective scan operation, and finally (6) a post processing unit (PPU) performing the computations after selective scan.

\begin{figure}[t] 
\centering
\includegraphics[width=0.43\textwidth]{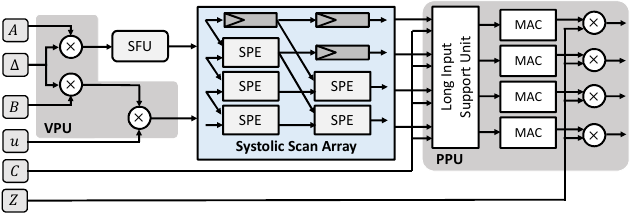}
\vspace{-0.5em}
\caption{Dataflow of selective SSM in \proposed.}
\label{fig:dataflow_selective_ssm}
\vspace{-1.0em}
\end{figure}

\textbf{Dataflow.} The execution flow of \proposed generally follows the computational sequence of an end-to-end Vision Mamba (\fig{fig:vim_architecture}) where the required compute units are activated on demand based on the operation. Below we focus on the dataflow of the selective SSM block using \fig{fig:dataflow_selective_ssm}.

At the start of execution, the required data is initially loaded into the on-chip buffer via DMA. The execution of selective SSM on \proposed involves not only the SSA but also the VPU, SFU, and PPU. First, the VPU conducts $\Delta \mathbf{A}$ and $\Delta \mathbf{B}\cdot u$ computations. Next, $\Delta \mathbf{A}$ is processed by the SFU to apply the exponential function. The computed $\Delta \mathbf{A}$ and $\Delta \mathbf{B}\cdot u$ are then processed by the SSA. The resulting output is subsequently handled by the PPU, where it undergoes a MAC operation with $\mathbf{C}$ using the MAC array, followed by a multiplication with $\mathbf{Z}$. The PPU also contains a Long Input Support Unit (LISU) that accumulates partial results, which we discuss in detail in \sect{sect:parallel_scan_array}.

\subsection{Systolic Scan Array (SSA) Design}
\label{sect:parallel_scan_array}

Despite various performance optimizations employed in GPU-based selective SSM, compute utilization remains low due to the inherent sequential dependencies of selective scan (\sect{sect:roofline_analysis}). We propose a systolic array~\cite{why_systolic, systolic_VLSI} architecture named Systolic Scan Array (SSA), tailored to the unique dataflow of Mamba's selective scan operation.

\begin{figure}[t] 
\centering
\includegraphics[width=0.45\textwidth]{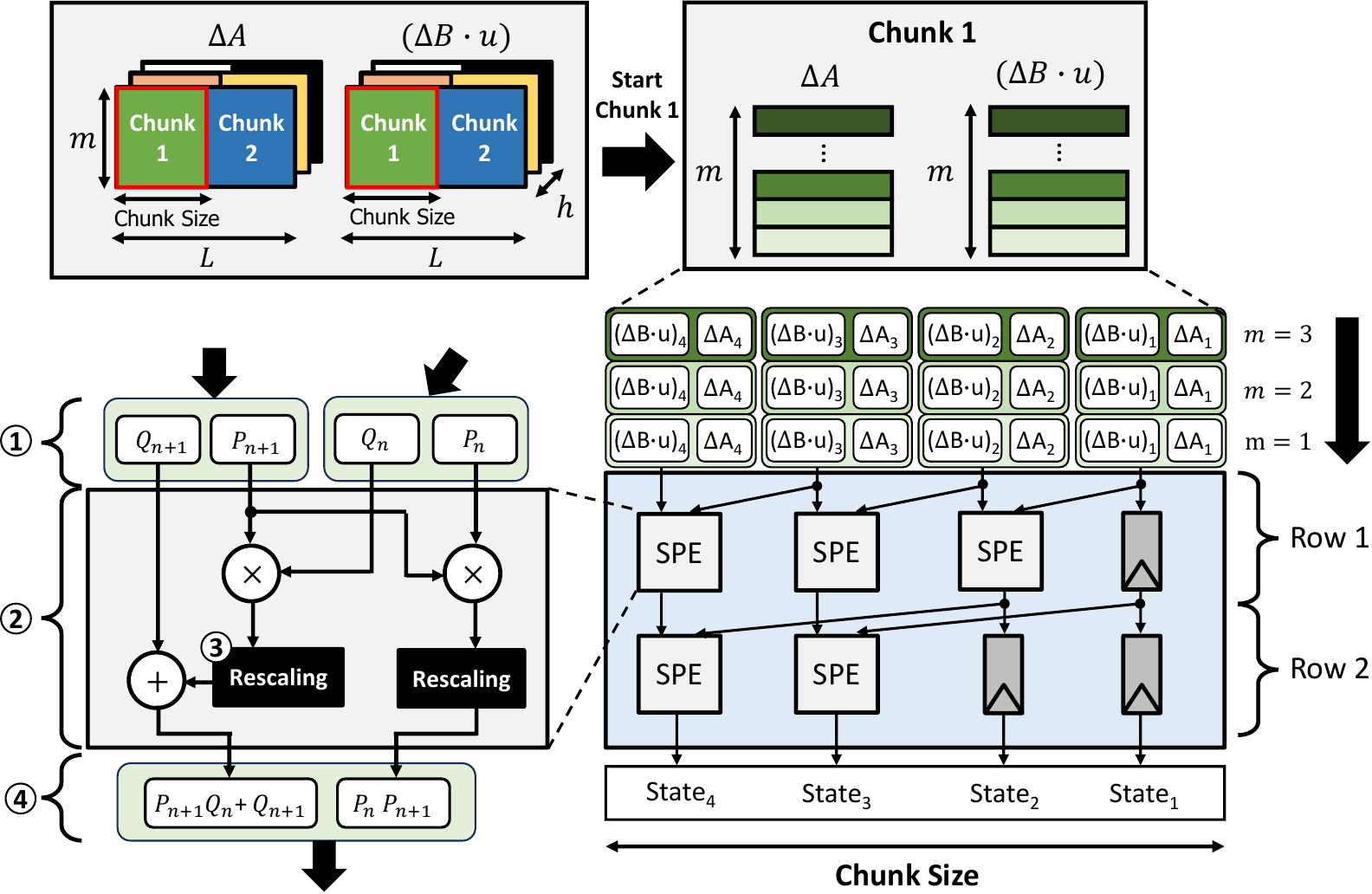}
\caption{Detailed microarchitecture of the SSA, the SPE, and our chunk-wise parallel scan dataflow.}
\label{fig:chunk_parallel_scan_array}
\vspace{-1.0em}
\end{figure}

{\bf Why systolic arrays?} Systolic array architectures are parallel processing systems designed to efficiently perform repetitive computational tasks by organizing Processing Elements (PEs) in a regular, grid-like structure. Adapting a target algorithm for systolic array-style implementation requires data to flow in a \emph{rhythmic, pipeline-like fashion between the PEs}, where each PE performs part of the computation before passing the data to the next PE. Our key observation is that the dataflow of the Kogge-Stone algorithm~\cite{kogge-stone} can be refactored into a systolic dataflow, as illustrated in \fig{fig:chunk_parallel_scan_array}, enabling low-latency \emph{local} inter-PE communication during selective scan. Recall from \sect{sect:roofline_analysis} that one of the primary performance bottlenecks of the baseline GPU system's selective scan operation lies in the limited on-chip SRAM storage, which limits the exploitation of parallelism during the scan operation and frequently spills intermediate results to off-chip memory. In our SSA architecture, the Scan Processing Elements (SPEs) communicate only with their neighboring SPEs, leading to high bandwidth and low-latency inter-SPE communication, obviating the need for allocating intermediate state variables in on-chip SRAM storage.

{\bf SSA dataflow.} We propose a \textit{chunk-wise} parallel scan dataflow, which is illustrated in \fig{fig:chunk_parallel_scan_array}. In our chunk-wise dataflow, $\Delta A$ and $\Delta B \cdot u$ are partitioned along $L$-dimension. Each pair of $\Delta A$ and $\Delta B \cdot u$ row vectors, corresponding to the same state dimension, is sequentially fed into the SSA.  The SSA consists of multiple SPEs and registers, where each row of SPEs receives the data from two input paths, and the registers store the intermediate results to efficiently maintain data dependencies. Each input path consists of two pairs of $P_n$ and $Q_n$ derived from $\Delta A$ and $\Delta B \cdot u$, respectively (Step \textcircled{1}). The SPE integrates two multipliers and one adder to execute the core operation of selective scan as shown in \fig{fig:MambaAccel_kogge_stone_implementation}(a). Unlike conventional MAC units that contain just a single multiplier, SPE is integrated with two multipliers to process $P_nP_{n+1}$ and $P_{n+1}Q_n+Q_{n+1}$ concurrently (Step \textcircled{2}), allowing it to directly propagate the output results to neighboring SPEs in a lockstep manner. During the SPE's execution, two rescaling operations are performed to accommodate quantized inputs (Step \textcircled{3}), and we detail its role in \sect{sect:quant}. The SPE produces a pair of outputs $P_nP_{n+1}$ and $P_{n+1}Q_n+Q_{n+1}$ (Step \textcircled{4}), which are subsequently propagated to downstream SPEs. To maintain accuracy, intermediate value $P_{n+1}Q_n+Q_{n+1}$ is computed using fixed-point representation with 2 extra fractional bits. If the SPE resides in the last row of SSA, the computed output $P_{n+1}Q_n+Q_{n+1}$ represents the final scan result.

\begin{figure}[t] 
\centering
\includegraphics[width=0.43\textwidth]{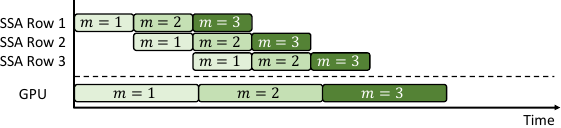}
\caption{Comparison of how the selective scan operation for the three rows of $\Delta A$ and $\Delta B \cdot u$ in \fig{fig:chunk_parallel_scan_array} is handled between the SSA and the GPU. Here, $m$ denotes the state dimension (\emph{m}).}
\label{fig:Pipeline_parallelism}
\vspace{-0.5em}
\end{figure}

The key advantages of the chunk-wise parallel dataflow are twofold. First, in a GPU-based selective scan, each thread within a warp performs a single computation per step and exchanges a partial result with other warps through on-chip memory. However, high-resolution image processing in Vision Mamba necessitates multiple iterative scan operations, resulting in substantial memory traffic that may overflow into off-chip memory. Thanks to the SSA's systolic array-based design, the SPEs can directly forward partial results to neighboring SPEs, eliminating the need for explicit memory accesses. This exploits output reuse from neighboring SPEs, reducing memory traffic and improving computational efficiency. Second, the current GPU-based selective SSM implementation sequentially conducts the scan operation across all state dimensions due to the use of a fused single kernel (i.e., Steps \ding{182}, \ding{183}, \ding{184}, and \ding{185} in \fig{fig:vim_architecture}(b)) which, as discussed in \sect{sect:roofline_analysis}, introduces bottlenecks that limit parallelism. In contrast, SSA eliminates this bottleneck by allowing different state dimensions to be processed independently and in parallel. Since each state dimension’s scan operation is independent, multiple scan operations can be executed across different rows in the SSA in a pipelined manner (\fig{fig:Pipeline_parallelism}),   significantly improving compute utilization and overall scan throughput. Furthermore, by deploying multiple SSA units in parallel, the scan workload can be partitioned into multiple chunks and processed concurrently, further enhancing parallel execution efficiency.

\begin{figure}[t] 
\centering
\includegraphics[width=0.45\textwidth]{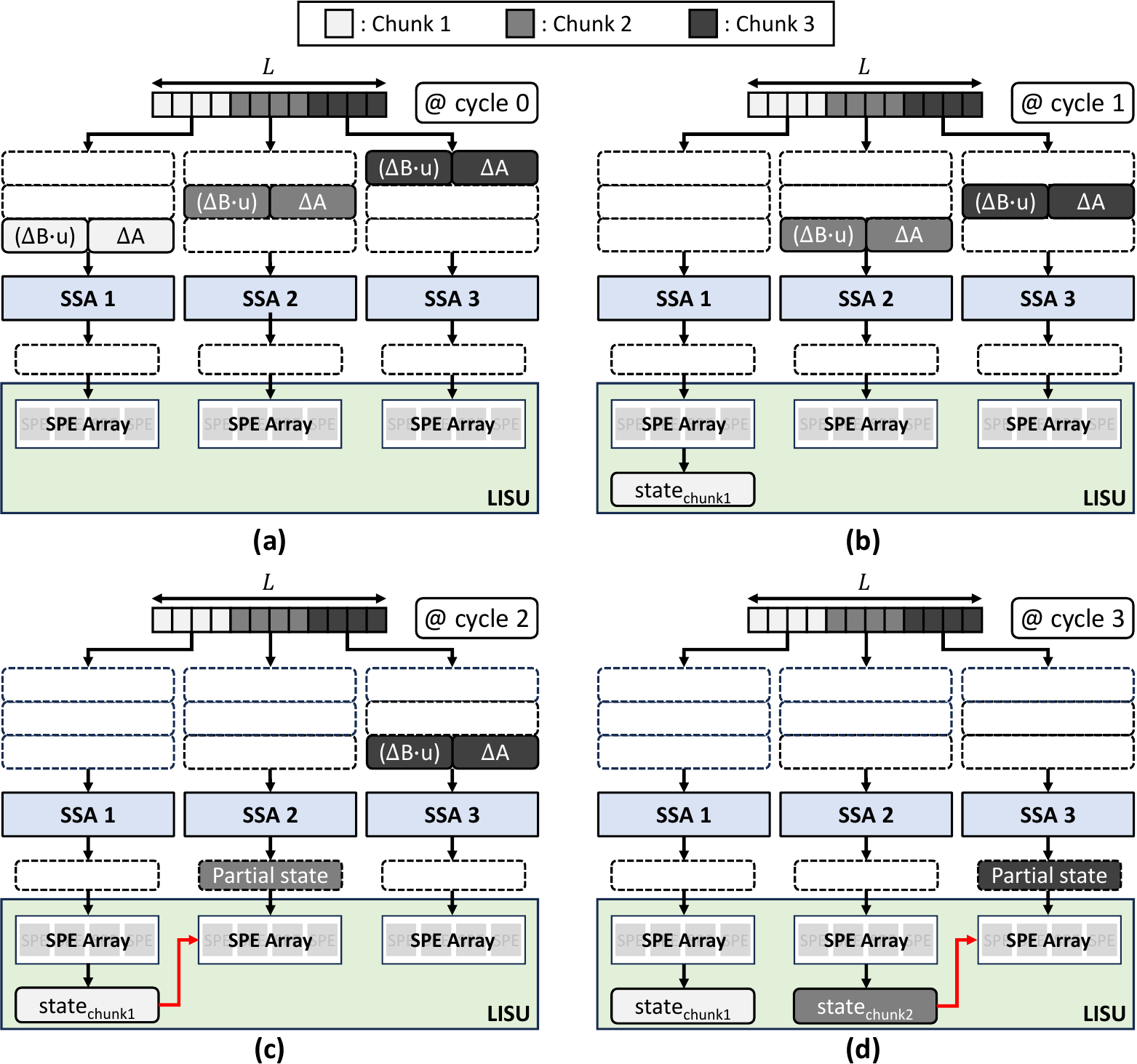}
\caption{Example showing how LISU resolves the inter-chunk dependencies across three chunks. }
\label{fig:MambaAccel_Long_Input_support_unit_vertical}
\vspace{-0.5em}
\end{figure}

\textbf{Remaining challenge of chunk-wise dataflow.} Despite the advantages of our chunk-wise dataflow, an important challenge remains due to how chunk partitioning is conducted along the $L$-dimension. Because the SSA's output includes only the partial state for its chunk, we introduce the Long Input Support Unit (LISU) as part of the PPU in order to efficiently handle inter-chunk dependencies (\fig{fig:MambaAccel_Long_Input_support_unit_vertical}). LISU consists of an additional row of SPEs designed to minimize on-chip buffer usage by enabling direct sharing of partial states across different chunks and eliminating the need for intermediate off-chip memory accesses.

\fig{fig:MambaAccel_Long_Input_support_unit_vertical} assumes a scenario where three SSAs are instantiated and the workloads are divided into three chunks, with each chunk assigned to a different SSA. The input chunks are fed into the arrays at one-cycle intervals to facilitate seamless execution and efficient scheduling. The following steps describe how LISU is integrated into the processing pipeline. (a) During cycle 0, chunks (i.e., $\Delta \mathbf{A}$ and $\Delta \mathbf{B} \cdot u$) are allocated to distinct SSAs. (b) In the following cycle 1, chunk 1 is processed by SSA 1, thereby completing its state computation ($state_{chunk1}$). (c) In cycle 2, chunk 2 is processed by SSA 2; however, the computation of chunk 2’s state requires the previous state value (i.e., the state of chunk 1, $state_{chunk1}$). Consequently, the LISU supplies the SPE array with $state_{chunk1}$ along with the partial state computed by SSA 2 as inputs. (d) Finally, in cycle 3, after computing $state_{chunk2}$, the LISU inputs both $state_{chunk2}$ and SSA 3’s output into the SPE array. Overall, LISU supports efficient inter-chunk communication, enhancing parallelism.

\subsection{Special Function Unit (SFU) Design}

\begin{figure}[t] 
\centering
\includegraphics[width=0.45\textwidth]{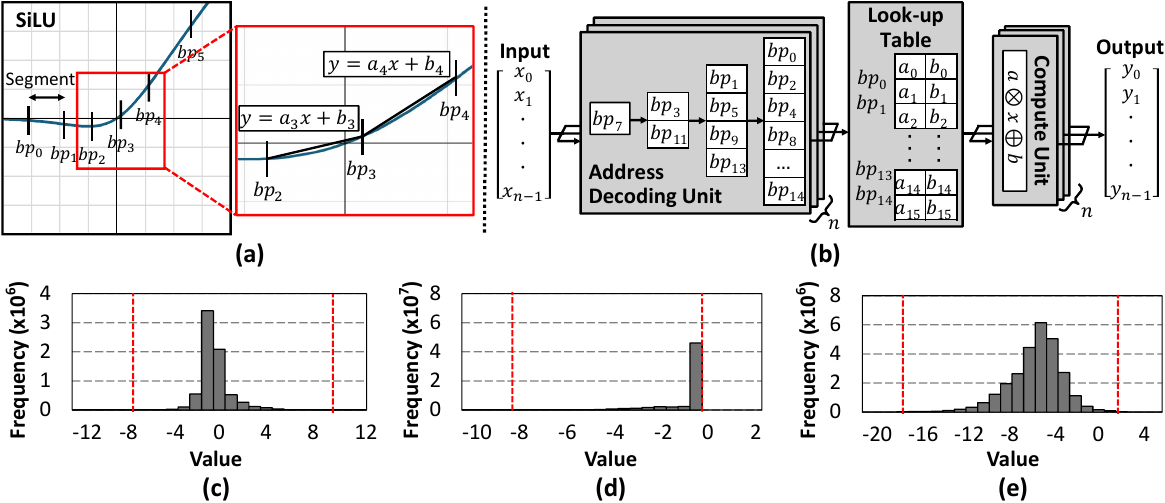}
\caption{(a) Example of how SiLU function can be approximated using linear interpolation. (b) \proposed's profile-guided SFU with a 16-entry LUT. The SFU locates the segment in which each value in the input vector $x$ is located, fetches the relevant coefficients from the LUT, and performs linear interpolation. (c), (d), and (e) show the input distributions for SiLU, exponential, and softplus functions, respectively, during Vision Mamba inference. Each graph has two red dashed lines, within which 99.9\% of the input data falls. }
\label{fig:MambaAccel_SFU_microarchitecture}
\end{figure}

We design the SFU to efficiently execute the non-linear activation functions used in Vision Mamba, specifically, the SiLU, exponential, and softplus operations, without incurring excessive area and energy overhead. Inspired by prior works \cite{nn_lut, flex_sfu}, our design adopts a lookup table (LUT)-based approach to approximate these functions. This significantly reduces arithmetic computations while preserving accuracy.

\textbf{Profile-guided refinement of LUT-based SFU.} As shown in \fig{fig:MambaAccel_SFU_microarchitecture}(a), we partition the target non-linear functions into non-uniform segments using breakpoints, or `$bp$'s. Each segment is then approximated with piecewise linear interpolation and represented as a linear function $(ax+b)$ between two breakpoints, with the precomputed coefficients $a$ and $b$ stored in LUTs. Depending on the number of segments used to approximate the nonlinear function, the number of LUT entries that hold coefficients \(a\) and \(b\) can vary. Our profiling results indicate that a 16-entry LUT is sufficient for the exponential function, and a 32-entry LUT maintains high accuracy for SiLU and softplus, demonstrating the feasibility of our low-cost approximation technique. We will evaluate the robustness of our LUT-based SFU design in \sect{sect:model_accuracy}.

To identify optimal breakpoints and coefficients, we utilize gradient descent, following the method described in~\cite{flex_sfu}, in combination with profile-guided refinement of the LUT. Specifically, we analyzed the input distributions of SiLU, exponential, and softplus functions during Vision Mamba model's inference with several sample images. As shown in ~\fig{fig:MambaAccel_SFU_microarchitecture}(c,d,e), approximately 99.9\% of inputs for the SiLU, exponential, and softplus functions fall within the range -8.7 to 10.2, -8.5 to 0, and -17.6 to 2.7, respectively. Using this information, we heuristically restrict the possible values that breakpoints can take during gradient descent within these input ranges for each target nonlinear function. This leads to higher accuracy in the region where most inputs occur, minimizing both approximation error and LUT size. 

\textbf{SFU architecture.}
As illustrated in~\fig{fig:MambaAccel_SFU_microarchitecture}(b), the SFU is made up of three submodules: the address decoding unit (ADU), the LUT, and the compute unit (CU). The ADU stores all breakpoint values $bp$ and uses binary search to determine the segment corresponding to each input. The LUT stores precomputed coefficients $a$ and $b$ for piecewise linear interpolation. Finally, the CU fetches relevant coefficients from the LUT and performs computations with the input to generate the final output. To improve throughput, the SFU processes vectorized inputs by incorporating multiple pairs of ADUs and CUs. Each ADU-CU pair processes one input value from the vectorized input independently. For example, the input \(x_0\) is first processed by the $0$-th ADU. After the appropriate coefficients are fetched from the LUT, they are then sent to the $0$-th CU for the final computation. The same process occurs for each input and the corresponding ADU-CU pair. Additionally, to enable concurrent coefficient fetching by multiple CUs from a single LUT, we employ a crossbar interconnect that connects the LUT to all CUs.

\begin{table}[t]
\centering
\caption{Accuracy comparison (Top-1 and Top-5) when applying tensor- or channel-granularity quantization to input activations.}
\label{tab:accuracy_result_with_granularity}
\footnotesize
\resizebox{\linewidth}{!}{
\setlength{\tabcolsep}{9pt} 
\renewcommand{\arraystretch}{0.95} 
\begin{tabular}{ccc}
\toprule
{\bf Quantization Type} & {\bf Top-1 Accuracy} & {\bf Top-5 Accuracy} \\
\midrule
Baseline & 76.04\% & 93.00\% \\
Tensor-granularity & 14.67\% & 30.00\% \\
Channel-granularity & 75.54\% & 92.74\% \\
\bottomrule
\end{tabular}
}
\end{table}

\subsection{Hybrid, Hardware-friendly (H2) Quantization}
\label{sect:quant}

GPU-based Vision Mamba utilizes NVIDIA's Automatic Mixed Precision (AMP)~\cite{AMP} inferencing feature to dynamically quantize weights and activations from FP32 to FP16, achieving higher performance while not sacrificing high accuracy. Despite these optimizations, as discussed in \sect{sect:roofline_analysis}, intermediate activation tensors in selective SSM blocks generate redundant off-chip memory traffic due to the limited on-chip SRAM capacity in edge devices. To address this inefficiency, we introduce our \emph{hybrid, hardware-friendly} (H2) quantization technique, tailored to the unique data distribution of Vision Mamba’s model and activations. Our approach is based on a lightweight post-training quantization (PTQ) method using a uniform integer quantization scheme. By quantizing both weights and activations to 8-bit integer precision (INT8), \proposed not only alleviates memory constraints in edge devices but also simplifies the design of adders and multipliers in the SPE (\fig{fig:chunk_parallel_scan_array}), enabling a more area-efficient accelerator design. This makes \proposed particularly well-suited for edge devices with stringent memory and computational constraints. We detail the two key components of our H2 quantization below.

\begin{figure}[t]
\centering
\includegraphics[width=0.48\textwidth]{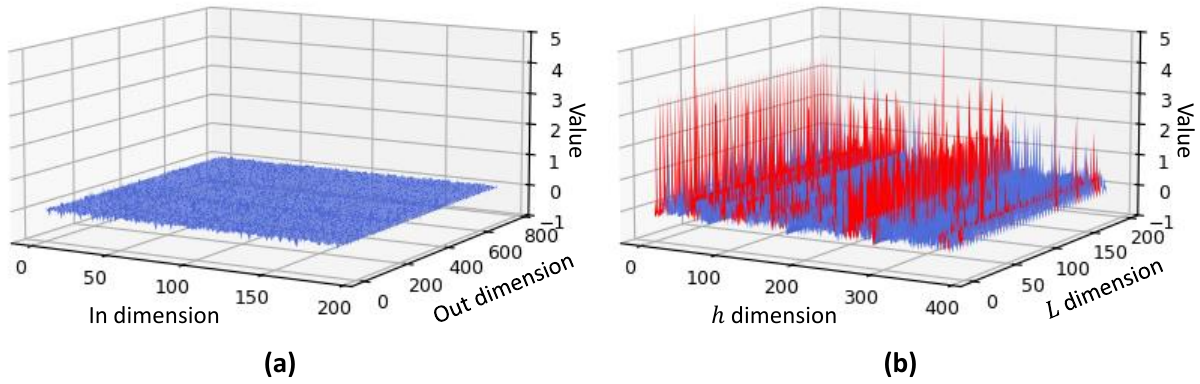}
\caption{Magnitude of (a) the weights in the input linear projection layer in the Vision Mamba Encoder (Step \textcircled{3} in \fig{fig:vim_architecture}(a)) and (b) input activation $u$ in the selective SSM block. For brevity, we only show the characterization results over the first Vision Mamba Encoder as other linear layers and activations in the selective SSM block exhibit similar trends.}
\label{fig:Value_distribution}
\vspace{-1.0em}
\end{figure}

\textbf{Hybrid quantization.} The simplest and most common quantization strategy applies a uniform scaling factor across the entire tensor, referred to as tensor-granularity quantization. However, when applied to the selective SSM block’s activations, this approach leads to a substantial drop in model accuracy (\tab{tab:accuracy_result_with_granularity}). As illustrated in \fig{fig:Value_distribution}, unlike the selective SSM block's model weights, which follow a relatively uniform distribution, the activations in our selective SSM exhibit high variance in their values. Specifically, we observe several outlier channels in the hidden dimension of the activation tensors, where global maximum and minimum values deviate significantly from those of other channels.

To address this issue, we employ a hybrid quantization approach. All linear layer's weights are quantized using tensor-granularity quantization, as they exhibit low variance in their value distribution. In contrast, channel-granularity quantization is applied to activations in selective SSM blocks to account for channel-wise distribution variations. Our hybrid quantization is implemented by determining scaling factors separately for weights and activations (Equation \eqref{eq:quant_equation} in \sect{sect:background_quantization}). Since model weights remain fixed at runtime, their scaling factors are precomputed before inference. In contrast, activation values vary dynamically at runtime, requiring a calibration step to estimate global maximum and minimum values from representative sample data. We observe that using only 1\% of the test dataset (500 randomly sampled images from the 50,000-image ImageNet-1K dataset) provides a robust estimation of global maximum and minimum values for configuring our scaling factors. Overall, our hybrid quantization strategy effectively mitigates accuracy loss, achieving an error margin below 1\%p, as detailed in \sect{sect:model_accuracy}.

\textbf{Hardware-friendly approximation of scaling factors.} While our hybrid quantization effectively reduces the memory usage of selective SSM blocks, it also introduces non-negligible computational overhead. Consider the execution of a selective SSM block using \proposed's SSA. An SPE receives four INT8-quantized inputs, $P_n$, $P_{n+1}$, $Q_n$, and $Q_{n+1}$:
\begin{equation}
    P_n=\lceil {{\Delta \mathbf{A}_n}\over{s_{\Delta  \mathbf{A}}}} \rfloor, \,\,\,
    P_{n+1}=\lceil {{\Delta  \mathbf{A}_{n+1}}\over{s_{\Delta  \mathbf{A}}}} \rfloor
\end{equation}
\begin{equation}
    Q_n=\lceil {{(\Delta  \mathbf{B}\cdot u)_n}\over{s_{\Delta  \mathbf{B}\cdot u}}} \rfloor, \,\,\,
    Q_{n+1}=\lceil{{(\Delta  \mathbf{B}\cdot u)_{n+1}}\over{s_{\Delta  \mathbf{B}\cdot u}}} \rfloor, \,\,\,
\end{equation}
where $s_{\Delta A}$ is the scaling factor of $P_n$, $P_{n+1}$, and $s_{\Delta  \mathbf{B} \cdot u}$ is the scaling factor of $Q_n$, $Q_{n+1}$, and $\Delta  \mathbf{A}_n$, $\Delta  \mathbf{A}_{n+1}$, $(\Delta  \mathbf{B}\cdot u)_n$, and $(\Delta  \mathbf{B}\cdot u)_{n+1}$ are the original floating point counterparts of $P_n$, $P_{n+1}$, $Q_n$, and $Q_{n+1}$, respectively. For an SPE to calculate the partial state, it must perform $P_{n+1}Q_n+Q_{n+1}$. The multiplication term is approximately expressed as follows:
\begin{equation}
    P_{n+1}Q_n\approx{{\Delta  \mathbf{A}_{n+1} \times (\Delta  \mathbf{B}\cdot u)_{n}}\over{s_{\Delta  \mathbf{A}} \times s_{\Delta  \mathbf{B} \cdot u}}}
\end{equation}
However, it is important to note that $P_{n+1}Q_n$ cannot be added directly to $Q_{n+1}$, as the scaling factors (denominators) of the two quantized terms differ. Therefore, $P_{n+1}Q_n$ must be rescaled by multiplying it with $s_{\Delta  \mathbf{A}}$ to match the scaling factors\cite{tender}.  Similarly, the other intermediate output of the SPE,  $P_nP_{n+1}$, must also undergo rescaling.

\begin{figure}[t] 
\centering
\includegraphics[width=0.45\textwidth]{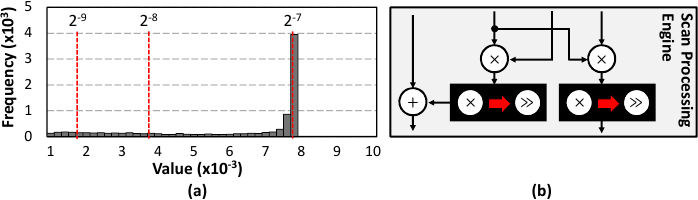}
\caption{(a) Histogram of $\Delta A$ scaling factor for backward SSM. (b) The rescaling process is changed from expensive multiplication to relatively cheaper shift operation with hardware-friendly scaling factor approximation.}
\label{fig:DeltaA_approximation}
\end{figure}

To reduce the hardware complexity associated with rescaling, we propose a hardware-friendly approximation of the scaling factors. By profiling samples from the ImageNet-1K dataset, we observe that most $s_{\Delta \mathbf{A}}$ values cluster near powers of two. As illustrated in \fig{fig:DeltaA_approximation}(a), most  $s_{\Delta  \mathbf{A}}$ values fall between $2^{-9}$ and $2^{-7}$, so we round them to the nearest power of two. This approximation allows us to replace the multiplication operation for rescaling with a shift operation, as shown in \fig{fig:DeltaA_approximation}(b), significantly improving hardware efficiency while preserving model accuracy.

\section{Methodology}
\label{sect:methodology}

\textbf{Performance.} 
We model \proposed as a cycle-level  simulator using C++. To compare the performance and energy consumption of \proposed with GPUs designed for edge environments, we use NVIDIA's Jetson AGX Xavier~\cite{JetsonAGXXavier} as the baseline GPU (denoted \emph{edge GPU}). This GPU is fabricated using TSMC's 12 nm process technology with a die size of 350 mm$^2$, equipped with 16 GB of LPDDR4X memory, and has a thermal design power (TDP) of 30W. \tab{tab:spec_comparison} compares the key system configurations of NVIDIA Jetson AGX Xavier and our \proposed architecture.

\textbf{Area.} We estimate \proposed's area by implementing it in RTL using SystemVerilog. The RTL is synthesized with the Synopsys Design Compiler, targeting an operating frequency of 1 GHz and using a 65 nm standard-cell library. The area of the on-chip scratchpad memory is estimated using CACTI 7.0~\cite{cacti}. Since CACTI only supports 32 nm technology, we scale down \proposed's synthesized 65 nm design to 32 nm following the methodology suggested in \cite{technology_scaling}.

\begin{table}
\centering
\caption{System configurations.}
\label{tab:spec_comparison}
\footnotesize
\begin{tabular}{ccc}
\toprule
\multirow{2}{*}{} & \bf NVIDIA & \multirow{2}{*}{\bf \proposed} \\
& \bf Jetson AGX Xavier & \\
\midrule
\multirow{3}{*}{Compute unit} & 512 CUDA Cores & \makecell{8 Systolic Scan Arrays \\ (16 chunk size)} \\
 & \makecell{64 Tensor Cores} & \makecell{1 GEMM Engine \\ (64$\times$64 PEs)} \\
\midrule
Operating frequency & 1.38 GHz & 1 GHz \\
\midrule
GEMM throughput & 11 TFLOPS & 8 TOPS \\
\midrule
On-chip memory & 512 KB & 384 KB \\
\midrule
\multirow{2}{*}{\makecell{Off-chip \\ memory bandwidth}} & \multirow{2}{*}{136.5 GB/s} & \multirow{2}{*}{136.5 GB/s} \\
 & & \\
\bottomrule
\end{tabular}
\vspace{-1em}
\end{table}

\begin{table}[t]
\centering
\caption{Vision Mamba model configurations.}
\label{tab:vision_mamba_config}
\scriptsize
\resizebox{\linewidth}{!}{
\begin{tabular}{ccccc}
\toprule
\multirow{2}{*}{} & \multirow{2}{*}{\makecell{\bf Model \\ \bf parameters}} & \multirow{2}{*}{\makecell{\bf\# Encoder \\ \bf blocks}} & \multirow{2}{*}{\makecell{\bf Hidden \\ \bf dimension}} & \multirow{2}{*}{\makecell{\bf State \\ \bf dimension}} \\
 &  &  &  &  \\
\midrule
Tiny  & 7 M  & 24  & 192  & 16 \\
Small & 26 M & 24  & 384  & 16 \\
Base  & 98 M & 24  & 768  & 16 \\
\bottomrule
\end{tabular}
}
\vspace{-1em}
\end{table}

\textbf{Energy.} The energy consumed by the processor's logic gates is derived by multiplying the total power consumption (both dynamic and static) of the RTL-synthesized \proposed design by its end-to-end inference time. The energy consumption of the on-chip buffers is estimated using CACTI, while the energy consumed by the off-chip memory is calculated by multiplying the number of bytes transferred across the off-chip interface by 4 pJ/bit, which corresponds to the LPDDR4 energy consumption per bit transfer~\cite{sohn2024hbmpim}.

\textbf{Model accuracy.} We evaluate \proposed on three Vision Mamba models of varying sizes (Tiny, Small, and Base), whose configurations are summarized in \tab{tab:vision_mamba_config}. When measuring model accuracy, we use the ImageNet-1K~\cite{imagenet} dataset, which contains 50,000 images at a resolution of 224$\times$224. To demonstrate the robustness of our proposed method, we compare the model accuracy of our proposal against an FP16-based baseline that employs PyTorch's Automatic Mixed Precision (AMP) feature~\cite{AMP}, which results in minimal accuracy degradation compared to the original FP32 model (an average of $0.04\%$ accuracy loss vs. FP32).

\section{Evaluation}

\begin{figure}[t] 
\centering
\includegraphics[width=0.475\textwidth]{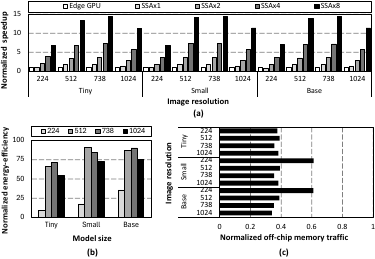}
\caption{The (a) speedup, (b) energy-efficiency, and (c) off-chip memory traffic during selective SSM block's execution with \proposed. All results are normalized to the edge GPU system.
}
\label{fig:Evaluation_selective_scan}
\end{figure}

\subsection{Performance and Energy-Efficiency}

\textbf{Selective SSM speedup and energy-efficiency.}
We evaluate \proposed's impact on the performance and energy efficiency of selective SSM across three dimensions: number of SSAs, image size, and model type. As shown in \fig{fig:Evaluation_selective_scan}(a), \proposed consistently outperforms the edge GPU with an average 11.6$\times$ speedup. This gain results from the chunk-wise dataflow, which exploits high parallelism. Additionally, as the number of SSAs increases, \proposed achieves scalable performance improvements and speedups, aided by the LISU within the PPU for efficient inter-SSA communication.

In terms of energy-efficiency, \proposed significantly reduces energy consumption compared to the edge GPU during selective SSM execution (\fig{fig:Evaluation_selective_scan}(b)). This improvement stems from three key factors. First, \proposed lowers energy consumption by reducing the selective SSM's latency. Second, our H2 quantization techniques enable the use of integer-based operations instead of floating-point operations, substantially decreasing energy consumption per math operation~\cite{horowitz20141, tpuv4i}. Finally, off-chip memory traffic is reduced by an average of 2.5$\times$ (\fig{fig:Evaluation_selective_scan}(c)), further contributing to \proposed's significantly higher energy-efficiency.

\begin{figure}[t] 
\centering
\includegraphics[width=0.485\textwidth]{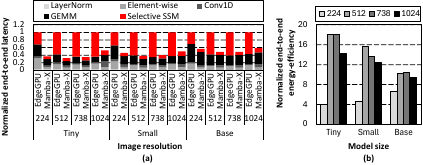}
\caption{The normalized end-to-end (a) latency breakdown and (b) energy-efficiency of \proposed compared to the edge GPU.}
\label{fig:Evaluation_end_to_end}
\vspace{-0.5em}
\end{figure}

\textbf{End-to-end speedup and energy-efficiency.}
We now examine how \proposed's improvement in selective SSM performance translates into end-to-end Vision Mamba speedup by breaking down the inference latency of \proposed and the edge GPU (\fig{fig:Evaluation_end_to_end}). \proposed noticeably reduces end-to-end latency compared to the edge GPU across all inference scenarios, though the extent of this reduction varies with input image size.
The significant latency reduction is primarily due to the decreased processing time of the selective SSM block (red bar in \fig{fig:Evaluation_end_to_end}(a)), while GEMM execution time remains comparable to that of the edge GPU (black bar). As the model size increases (e.g., Base), the magnitude of latency reduction diminishes because the performance is increasingly dominated by the larger proportion of GEMM operations.
Nonetheless, by substantially reducing the latency of selective SSM—the most critical performance bottleneck in Vision Mamba—\proposed achieves an average 2.3$\times$ end-to-end speedup. Notably, this speedup is achieved despite using only a small fraction of the Jetson AGX Xavier's die area, a detail we elaborate on in the next subsection.

\subsection{\proposed's Area-Efficiency vs. Edge GPU}

This work focuses on deploying Vision Mamba on resource-limited edge devices, making high area efficiency one of the most critical design objectives of \proposed. \tab{tab:area_power_result} provides an area breakdown of \proposed under 32 nm and 12 nm technologies. When scaled to 12 nm, the total area of \proposed, configured with 8 SSAs and a GEMM engine featuring 64$\times$64 PEs, amounts to just 1.34 mm$^2$—only 0.4\% of the Jetson AGX Xavier's die size (350 mm$^2$ at 12 nm). Notably, SSAs occupy about 3\% of \proposed's total area, as their compute hardware supports only low-precision operations via our H2 quantization with a lightweight interconnect using short, local inter-SPE links, thanks to our systolic array based design. This substantial improvement in area efficiency enables \proposed to achieve a 601$\times$ increase in average end-to-end performance per unit area relative to the edge GPU.

\begin{table}[t]
\centering
\caption{Area breakdown (mm$^2$).}
\label{tab:area_power_result}
\small
\resizebox{\linewidth}{!}{
\setlength{\tabcolsep}{3pt} 
\renewcommand{\arraystretch}{0.95} 
\begin{tabular}{ccccccccc}
\toprule
& \bf SSA & \bf SFU & \bf VPU & \bf PPU &  \makecell{\bf GEMM \\ \bf Engine} & \makecell{\bf On-chip \\ \bf Buffer} & \bf Others & \bf Total \\
\midrule
\bf 32 nm  & 0.28 & 1.00 & 0.23 & 0.85 & 5.34 & 1.74 & 0.04 & 9.48 \\
\bf 12 nm  & 0.04 & 0.14 & 0.03 & 0.12 & 0.75 & 0.25 & 0.01 & 1.34 \\
\bottomrule
\end{tabular}
}
\end{table}

\subsection{\proposed's Effect on Model Accuracy}
\label{sect:model_accuracy}

\tab{tab:quant_result} summarizes the top-1 and top-5 accuracy results of the baseline and our proposed method. The baseline achieves a slightly higher accuracy, as all operations are performed using the floating point format with  higher precision than \proposed's INT8 based selective SSM. Our proposed method exhibits comparable accuracy for the Tiny, Small, and Base models, with only 0.75\%p, 0.59\%p, and 0.89\%p top-1 accuracy losses relative to baseline, respectively. These results indicate that our proposed method  incurs less than 1\%p accuracy degradation while significantly improving computational efficiency, making it a practical solution for Vision Mamba in resource-constrained environments.

\begin{table}
\centering
\caption{Accuracy results of the baseline and our proposed method.}
\label{tab:quant_result}
\small
\resizebox{\linewidth}{!}{
\setlength{\tabcolsep}{3pt} 
\renewcommand{\arraystretch}{0.95} 
\begin{tabular}{ccccccc}
\toprule
\multirow{2}{*}{\bf Quantization} & \multicolumn{2}{c}{\bf Tiny} & \multicolumn{2}{c}{\bf Small} & \multicolumn{2}{c}{\bf Base} \\
\cmidrule(lr){2-3} \cmidrule(lr){4-5} \cmidrule(lr){6-7}
 & \bf Top-1 & \bf Top-5 & \bf Top-1 & \bf Top-5 & \bf Top-1 & \bf Top-5 \\
\midrule
Baseline & 76.04\% & 93.00\% & 80.45\% & 95.08\% & 81.79\% & 95.64\% \\
Proposed & 75.29\% & 92.48\% & 79.86\% & 94.79\% & 80.90\% & 95.38\% \\
\bottomrule
\end{tabular}
}
\end{table}

\begin{figure}[t] 
\centering
\includegraphics[width=0.485\textwidth]{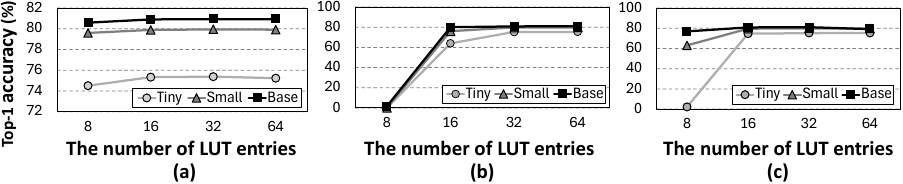}
\caption{Changes in accuracy as the number of LUT entries is varied for the (a) exponential, (b) SiLU, and (c) softplus functions.}
\label{fig:eval_lut_accuracy}
\vspace{-0.5em}
\end{figure}

\textbf{Sensitivity to SFU's LUT entries.}
\proposed's LUT-based SFU approximates non-linear functions using piecewise linear interpolation, where accuracy directly depends on the number of LUT entries. Therefore, selecting an optimal number of LUT entries that balances accuracy and hardware efficiency is crucial. \fig{fig:eval_lut_accuracy} illustrates how accuracy changes as the number of LUT entries varies for the exponential, SiLU, and softplus functions. \proposed uses 16 LUT entries for the exponential function and 32 LUT entries for the SiLU and softplus functions, as these configurations provide the best trade-off between accuracy and hardware efficiency.

\begin{figure}[t] 
\centering
\includegraphics[width=0.45\textwidth]{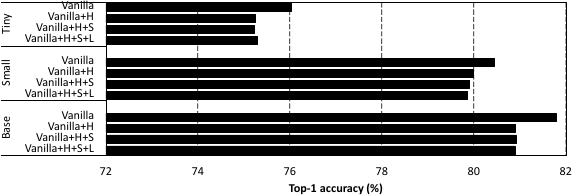}
\caption{Top-1 accuracy when the baseline Vision Mamba model (Vanilla) is applied with hybrid quantization (H), hardware-friendly approximation of scaling factors (S), and LUT-based SFU (L).}
\label{fig:Accuracy_result_with_quant_methods}
\vspace{-0.5em}
\end{figure}

\textbf{Ablation study.} \fig{fig:Accuracy_result_with_quant_methods} shows the accuracy impact of \proposed's quantization and approximation methods: hybrid quantization (H), hardware-friendly scaling factor approximation (S), and LUT-based SFU (L). Each method is applied incrementally, and its accuracy degradation is measured. Hybrid quantization causes the largest drop due to its scaling factors being precomputed from a limited calibration set from ImageNet-1K, leading to potential mismatches in real inference scenarios. In contrast, the impact of scaling factor approximation and LUT-based SFU is minimal.

\section{Related Work}

\textbf{Accelerators for SSM.}
Domain-specific accelerators for SSM have gained attention with the rise of modern sequence models like H3~\cite{H3, study_H3} and Mamba~\cite{mamba, vmamba, graph_mamba, graph_mamba_long_range, parameterization_SSM, simba, stg_mamba, pointmamba}. VGA~\cite{VGA_H3} introduces a specialized compute unit to accelerate Fast Fourier Transform (FFT) in H3. Unlike H3, Mamba-based vision models rely on selective SSM rather than FFT, which is the focus of our work. MARCA~\cite{marca} targets Mamba-based large language models (LLMs), proposing a reconfigurable architecture employing large on-chip SRAM and HBM for large-scale datacenter deployment of LLMs. In contrast, \proposed focuses on vision models with limited on-chip SRAM and LPDDR-based off-chip memory for edge environments.

\textbf{Quantization.}
Prior works~\cite{PTQ_ViT, PTQ4ViT, Repq_vit} propose PTQ schemes for ViTs that address challenges in executing the attention mechanism while minimizing accuracy loss. \cite{PTQ_ViT} introduces a calibration method to reduce quantization errors for the wide dynamic range of activations. PTQ4ViT~\cite{PTQ4ViT} enhances PTQ with a block-wise reconstruction approach, achieving a better balance of efficiency and accuracy through layer-wise optimization and mixed-precision quantization. RepQ-ViT~\cite{Repq_vit} applies scale reparameterization to reduce sensitivity to quantization noise. In contrast, our work focuses on the selective SSM of Vision Mamba, which has different operational characteristics from the attention mechanism.

\textbf{Accelerators for ViT.}
With the advent of ViT~\cite{vision_transformer}, several accelerators leveraging algorithm-hardware co-design have emerged. ViTCoD~\cite{vitcod} prunes and polarizes attention maps to address the self-attention bottleneck. ViTALiTy~\cite{vitality} exploits low-rank properties of linear attention to reduce computational and memory costs. HeatViT~\cite{heatvit} uses image-adaptive token pruning and 8-bit quantization on FPGAs for ViT inference. In contrast, our work focuses on accelerating Vision Mamba inference by targeting the selective SSM operation, a distinct challenge from reducing attention mechanism overhead in ViTs.

\section{Conclusion}

\proposed addresses Vision Mamba's computational and memory challenges using a Systolic Scan Array based on the Kogge-Stone algorithm, maximizing parallelism in selective scan operations while minimizing on-chip buffer usage and off-chip memory traffic. Additionally, \proposed uses a hybrid, hardware-friendly quantization scheme to reduce memory overhead while preserving accuracy. Compared to the baseline GPU, \proposed achieves 11.6$\times$ higher selective scan throughput, with an 11.5$\times$ energy efficiency boost and 601$\times$ better performance per area.

\bibliographystyle{ACM-Reference-Format}
\bibliography{main}

\end{document}